\documentstyle[prl,aps]{revtex}

\begin{document}

\title{
Landauer Conductance and Nonequilibrium Noise
of One-Dimensional Interacting Electron Systems
}

\title
{
Landauer Conductance and Nonequilibrium Noise of \\
One-Dimensional Interacting Electron Systems
}

\author
{
Akira {\sc Shimizu}\footnote{
E-mail address: shmz@ASone.c.u-tokyo.ac.jp
}}

\address
{
Institute of Physics,
University of Tokyo, Komaba \\
3-8-1 Komaba, Meguro-ku, Tokyo 153, Japan
}

\maketitle

\begin{abstract}
The conductance of one-dimensional interacting electron systems
is calculated in a manner similar to Landauer's argument for
non-interacting systems. Unlike in previous studies in which
the Kubo formula was used, the conductance is directly evaluated as
the ratio of current $J$ to the chemical potential difference
$\Delta \mu$ between right-going and left-going particles.
It is shown that both $J$ and $\Delta \mu$ are renormalized
by electron-electron (e-e) interactions, but their ratio, the conductance,
is not renormalized at all
if the e-e interactions are the only scattering mechanism.
It is also shown that nonequilibrium current fluctuation
at low frequency is absent in such a case.
These conclusions are drawn for both Fermi liquids (in which
quasi-particles are accompanied with the backflow)
and Tomonaga-Luttinger liquids.
\end{abstract}

\date{Received March 11, 1996}

\section{Introduction}
\label{intro}
The two-terminal conductance $G$ of
one-dimensional (1D) electron systems
without any scattering is given by the
Landauer formula\cite{landauer}
\begin{equation}
G = s/2 \pi
\label{LF}
\end{equation}
where $s$ denotes the degeneracy (which is usually
2 due to the spin degeneracy), and we take $e^2=\hbar=1$ throughout this paper.
When electron-electron (e-e) interactions are introduced,
the electron system behaves as
a so-called Tomonaga-Luttinger (TL)
liquid.\cite{TL1,TL2,TL3,TL4,TL5,early1,early2,early3,early4}
Early theoretical studies\cite{early1,early2,early3,early4} indicated that
the conductance of a TL liquid would be
\begin{equation}
G =  \gamma  s / 2 \pi,
\label{wrong}
\end{equation}
where $\gamma$ is a constant which depends on
the e-e interactions.\cite{early1,early2,early3,early4}
However, the result of a recent experiment\cite{tarucha} agrees with
eq.\ (\ref{LF}) rather than eq.\ (\ref{wrong}).

To explain this discrepancy,
several theoretical studies
based on the Kubo formula\cite{kubo} were reported.\cite{maslov,pono,kawabata}
However, they depend on specific models\cite{maslov,pono,kawabata}
and/or specific approximations.\cite{kawabata}
Moreover, although Kawabata\cite{kawabata} stressed that one must
carefully distinguish between the external and internal
fields, this point was not clear in most
work.\cite{early1,early2,early3,early4,maslov,pono}

In this paper, we show that the formula (\ref{LF}) holds
quite generally for 1D systems at low temperatures
if the e-e interactions are the only scattering mechanism.
This is explicitly shown for
{\it both} Fermi liquids and TL liquids.
The result is independent of details of models,
such as values of the Landau parameters or strength of e-e interactions.
Instead of using the Kubo formula, we calculate
$G$ directly by dividing the current by the chemical potential
difference.
This eliminates the above-mentioned difficulty
caused by the difference between the external and internal fields.

We also show that the nonequilibrium noise vanishes when
$G$ is given by eq.\ (\ref{LF}), irrespective of e-e interactions.

Throughout this paper,
it is assumed that the structure of the 1D system is smooth enough
and the temperature is low
enough, so that scattering by impurities, defects or phonons
is negligible. That is, we
treat ``clean" 1D systems in which e-e interactions
are the only scattering mechanism.
We consider the case $s=1$ because the generalization to
cases of $s \geq 2$ is trivial.
We assume zero temperature for simplicity.

\section{Nominal Conductance of 1D Fermi Liquids}
\label{GF}

It is generally believed that a
1D interacting electron system cannot be a Fermi liquid.
However,
it is very instructive to evaluate the Landauer conductance for
a 1D Fermi liquid because
(i) it provides considerable insight into
the underlying physics, and
(ii) real systems have finite length and finite intersubband excitation
energies (from the occupied subband, which is considered as the 1D system, to
an empty subband) and thus some real systems
might be well described as a Fermi liquid.
Note that the validity of eq.\ (\ref{LF}) for Fermi liquids
is {\it never trivial} because we need to consider the backflow\cite{noz} and
we cannot simply repeat Landauer's discussion\cite{landauer} with
the free electron mass replaced by the renormalized mass.

A Fermi liquid is characterized by the quasi-particle
distribution $n(k)$ and the Fermi wavenumber $k_F$.\cite{noz}
At equilibrium,
$n(k) = n^0(k) \equiv \Theta (|k| \leq k_F)$,
where $\Theta$ is a unit step function which is unity if the
argument is true and zero otherwise.
Under a small external perturbation,
$n$ would change slightly as
\begin{equation}
n(k) = n^0(k) + \delta n(k),
\label{n}
\end{equation}
and we may expand the corresponding change of the total energy $E$ as
\begin{equation}
\delta E =
\sum_k \epsilon^0_k \delta n(k) +
{1 \over 2L} \sum_k \sum_{k'} f(k,k') \delta n(k) \delta n(k'),
\end{equation}
where $L$ is the normalization length (system size).\cite{f}
By definition, we can take $f$ to be symmetric;
\begin{equation}
f(k,k')=f(k',k).
\end{equation}
The quasi-particle energy $\epsilon_k$ is the functional derivative of $E$:
\begin{equation}
\epsilon_k \{ n \}
=
\epsilon^0_k +
{1 \over L} \sum_{k'} f(k,k') \delta n(k'),
\label{eps}
\end{equation}
where by $\{ n \}$ we indicate that $\epsilon_k$ is
a functional of $n(k)$'s.

Let us evaluate the current $J$ when $n(k)$ is of the
form of the ``shifted Fermi distribution";
\begin{equation}
n(k) = \Theta(|k-q| \leq k_F),
\label{sfd}
\end{equation}
where $q$ is a small ($|q| \ll k_F$) wavenumber.
We will argue in the next section that this shifted-Fermi state
corresponds to the experimental situation
under the assumptions that
the sample is well fabricated and the temperature is low
enough.

We note that the above distribution is the same as what we
would find if we observed the equilibrium distribution $n^0(k)$
from a moving frame. The current for such a case
was calculated in \S1.2c of ref.\ \cite{noz}.
The current carried by a quasi-particle of momentum $k$ is
given by
\begin{eqnarray}
j_k &=&
{1 \over L} \left[ v_k - {1 \over L} \sum_{k'} f(k,k')
{\delta n(k') \over \delta  q} \right]
\\
&\simeq&
{v_k \over L} + {1 \over 2 \pi L} [f(k,k_F)-f(k,-k_F)],
\label{j}
\end{eqnarray}
where $v_k = \partial \epsilon_k^0 / \partial k$ is
the velocity of the quasi-particle, and we have used,
in the second line, the
fact that $\delta n$ is nonzero only near the Fermi points $k=\pm k_F$.
The total current is thus given by
\begin{eqnarray}
J &=& \sum_k \delta n(k) j_k \\
&\simeq&
{q \over \pi}[v_F + {1 \over 2 \pi }(f_{++}-f_{+-})],
\label{J}
\end{eqnarray}
where $f_{++} \equiv f(k_F,k_F)=f(-k_F,-k_F)$ and
$f_{+-} \equiv f(k_F,-k_F)=f(-k_F,k_F)$.
For a parabolic $\epsilon^0_k$, for example,
we have $v_k=  k / m^*$ and $v_F =  k_F / m^*$,
with $m^*$ being the renormalized mass, and
eqs.\  (\ref{j}) and (\ref{J}) show that
the current is the sum of this trivial contribution and
the backflow term.
At first sight, the backflow may appear to modify
the Landauer formula.
However, this is not the case, as we will now show.

Suppose that we add a right-going ($k>0$) quasi-particle
of minimum allowable energy. The energy cost defines
the chemical potential $\mu_+$ of right-going quasi-particles.
We can also define $\mu_-$
for left-going ($k<0$) quasi-particles.
From eqs.\ (\ref{n}), (\ref{eps}) and (\ref{sfd}) we have
\begin{eqnarray}
\mu_\pm
&=&
\epsilon_{\pm k_F+q} \{ n^0 + \delta n \} \\
&\simeq&
\epsilon^0_{\pm k_F + q} +
{1 \over L} \sum_{k'} f(\pm k_F, k') \delta n(k') \\
&\simeq&
\epsilon^0_{k_F} \pm v_F  q
+ {q \over 2 \pi} (-f_{\pm -} + f_{\pm +}).
\end{eqnarray}
Hence the chemical potential difference
\begin{eqnarray}
\Delta \mu &\equiv& \mu_+ - \mu_- \\
&\simeq&
2  q [v_F + {1 \over 2 \pi } (f_{++} - f_{+-})].
\label{dm}
\end{eqnarray}
By dividing $J$ by $\Delta \mu$, we obtain a conductance $G$,
which we here call the ``nominal conductance"
because the above discussions did not consider structures of
real samples.
We will show in the next section that $G$ equals
the observed conductance $G_{obs}$.

For the nominal conductance, eqs.\
(\ref{J}) and (\ref{dm}) yield
\begin{equation}
G \equiv J/\Delta \mu = 1/2 \pi .
\end{equation}
We thus find that $G$ is {\it not} renormalized by e-e interactions
because $J$ and $\mu$ are {\it both} renormalized {\it by the same amount}.

\section{1D Fermi Liquid Connected to Reservoirs}
\label{Fobs}

Our next task  is to show that $G_{obs}=G$.
In the above discussions we did not consider the ends of the 1D system.
Both ends of a real sample, however, are connected to
two- or three-dimensional systems, which are
called ``contacts" or ``reservoirs."\cite{landauer}
The left (L) and right (R) reservoirs have different
chemical potentials, $\mu_L$ and $\mu_R$, respectively.
The reservoirs are large enough, so that
they remain at equilibrium even in the presence of
a finite current $J$ between the reservoirs through the 1D system.

To avoid undesirable reflections at the boundaries between
the 1D system and the reservoirs, real samples are usually
fabricated in such a way that the
boundary regions are
smooth and long ($>$ Fermi wavelength).
We assume that this is the case.
Then, the wavefunction of a quasi-particle
propagating in the 1D system will
diffuse adiabatically into a reservoir, without suffering
any reflection.
By considering the time-reversed state of this process, we can also
deduce that there exists a set of excitation modes in reservoirs,
each of which evolves into a quasi-particle state in the 1D system.
Through these modes, reservoir L (R) provides the 1D system with
right-going (left-going) quasi-particles,
of energy less than $\mu_L$ ($\mu_R$).
Therefore, the reservoirs tend to drive the 1D system to
the state with the shifted Fermi distribution with
$\mu_+ = \mu_L$ and $\mu_- = \mu_R$.
This shifted Fermi state will be reached if it
is an approximate eigenstate of the 1D system.
This is indeed the case if the 1D system
is of infinite length because the shifted Fermi state
is the ground state in a moving frame, and hence is
an eigenstate in the rest frame.
(The infinite 1D system has translational invariance
because we have assumed the
absence of scattering by impurities, defects or phonons.)
For a finite-length system
such a state should also be an approximate eigenstate if
the length is long enough.
(In \S\ref{TLobs} we will present another line of reasoning,
where we will show
that other states cannot be realized as a steady state.)
We thus find that $G_{obs} \equiv J/(\mu_L-\mu_R)$
$= J/(\mu_+-\mu_-) = G$.
Therefore the observed conductance is {\it not} renormalized either:
$G_{obs}=1/2 \pi$.

\section{Nominal Conductance of TL Liquids}
\label{GTL}

We now evaluate the nominal conductance of clean TL liquids.
Since we are interested in the conductance in the limit of
low frequency and long wavelength,
only low-energy excitations are relevant.
The low-energy dynamics of TL liquids
is described by a fixed-point Hamiltonian,
which is diagonalized as\cite{TL1,TL2,TL3,TL4,TL5}
\begin{equation}
H = \sum_p \omega_p \hat b_p^\dagger \hat b_p
+ {\pi \over 2 L}
[v_N(\hat N_+ + \hat N_-)^2 + v_J (\hat N_+ - \hat N_-)^2],
\label{H}
\end{equation}
where $\hat b_p^\dagger$ ($\hat b_p$) is a boson creation
(annihilation) operator,
and $\hat N_+$ and $\hat N_-$ are the number operators which correspond to
the numbers of right-going and left-going particles, respectively.
Note that the e-e interactions renormalize the two velocities $v_N$ and
$v_J$ differently: $v_N \neq v_J$ for interacting systems, while
$v_N = v_J$ for non-interacting systems.\cite{TL1,TL2,TL3,TL4,TL5}
Let $N_p$ and $N_\pm$ be the eigenvalues of $\hat b_p^\dagger \hat b_p$
and $\hat N_\pm$, respectively.
Low-energy eigenstates of a TL liquid are completely labeled by
these quantum numbers,\cite{TL1,TL2,TL3,TL4,TL5}
and the eigenenergies are given by
\begin{equation}
E = \sum_p \omega_p N_p
+ {\pi \over 2 L}
[v_N(N_+ + N_-)^2 + v_J (N_+ - N_-)^2].
\label{ETL}
\end{equation}
The dc current
averaged over the sample length $L$, which
corresponds to the current component at
low frequency and long wavelength, is given by\cite{TL1,TL2,TL3,TL4,TL5}
\begin{equation}
J = v_J (N_+ - N_-)/L.
\label{JTL}
\end{equation}

Let us consider states with $N_p =0$ for all $p$ and
$N_+ \neq N_-$.
We will argue in the next section that such a state (with
$N_+ + N_- = 0$) is realized in real samples.
Let us define $\mu_\pm$ as in \S\ref{GF}.
We then have
\begin{equation}
\mu_\pm = {\partial E \over \partial N_\pm}
= {\pi \over L}[v_N(N_+ + N_-) \pm v_J(N_+ - N_-)],
\end{equation}
which shows that $\mu_\pm$ depends on {\it both} $v_N$ and $v_J$,
and is renormalized by e-e interactions.
However, the $v_N$ dependence disappears
if we take the difference;
\begin{eqnarray}
\Delta \mu &\equiv& \mu_+ - \mu_- \\
&=& 2 \pi v_J (N_+ - N_-)/L,
\end{eqnarray}
which is renormalized {\it only through $v_J$,
just as $J$ is}.
Consequently, the nominal conductance
is {\it not} renormalized at all by e-e interactions:
\begin{equation}
G \equiv J/\Delta \mu = 1/2 \pi.
\end{equation}

\section{TL Liquid Connected to Reservoirs}
\label{TLobs}

Let us explore the relationship between
the nominal conductance $G$ and the observed conductance $G_{obs}$.
As in \S\ref{Fobs}, we assume that both ends of the 1D system are connected to
reservoirs of chemical potentials $\mu_L$ and $\mu_R$,
and assume that the boundary regions are smooth and long.
Note that
the argument in \S\ref{Fobs} cannot be applied directly to
TL liquids because of the lack of
single-particle excitations.\cite{TL1,TL2,TL3,TL4,TL5}
Instead of modifying the argument, we take a different approach.
We argue that in the linear response regime
the steady state must be the state with the minimum energy
among states which satisfy given external conditions.
Otherwise, the system would be unstable and
would evolve into a state with lower energy.
For our purpose, it is convenient to
take the value of $J$ as the given external condition.
(This is just a Legendre transformation of a
theory in which $\Delta \mu$ is given.)
It is then clear from eqs.\ (\ref{ETL}) and (\ref{JTL}) that
the state with $N_p=0$ and $N_+ + N_- = 0$ has
the minimum energy for a given $J$.
Hence such a state must be realized as a steady state.
We further argue that
if $\mu_{L,R}$ were not equal to $\mu_{+,-}$ then
the system would be unstable because an
extra flow could be induced between the 1D system and the reservoirs.
Therefore, what is realized as the steady state is that
with $N_p=0$ for all $p$, $N_+ + N_- =0$ and $\mu_{L,R}=\mu_{+,-}$.
We thus find that $G_{obs} \equiv J/(\mu_L-\mu_R)$
$= J/(\mu_+-\mu_-) = G$.
Therefore the observed conductance is {\it not} renormalized either,
$G_{obs}=1/2 \pi$, in agreement with experiment.\cite{tarucha}

\section{Nonequilibrium Current Fluctuation}
\label{necf}

So far, we have considered the linear conductance.
The well-known fluctuation-dissipation theorem\cite{kubo}
states that the linear conductance
is related to the current fluctuation $\langle \delta J^2 \rangle$
{\it evaluated at equilibrium}, {\it i.e.}, at $J=0$.
Its zero-frequency component
vanishes at zero temperature:
\begin{equation}
\langle \delta J^2 \rangle_{J=0}^{\omega=0} = 0.
\label{EN}
\end{equation}
This result is known to generally hold because
it simply states that the power spectrum of the zero-point fluctuations
vanishes at $\omega=0$ (because the energy of a zero-point fluctuation
is proportional to $\omega$).
In the presence of nonzero $J$, on the other hand,
finite $\langle \delta J^2 \rangle_{J \neq 0}^{\omega=0}$ is generally
observed even at low temperatures.\cite{NENth1,NENth2,NENex1,NENex2}
This excess noise is called the nonequilibrium noise (NEN).

The NEN can vanish in some cases,
such as the case of
free 1D electrons without any scattering\cite{NENth1,NENth2}
or the case of electrons with strong dissipation.\cite{su,isqm,us}
Interestingly, we can show that
the NEN vanishes also for the 1D interacting
systems discussed in the previous sections;
\begin{equation}
\langle \delta J^2 \rangle_{J \neq 0}^{\omega=0} =0.
\end{equation}
To show this, we note that $\langle \delta J^2 \rangle$
is invariant under Galilean transformations.
Hence, we can calculate it in a moving frame.
For each of the steady states discussed in
\S\ref{GF} and \S\ref{GTL},
we can find a moving frame in which the state becomes the ground state.
For the ground state, $J=0$ and eq.\ (\ref{EN}) yields
$\langle \delta J^2 \rangle =0$.
Therefore, it should also be zero in the rest frame.\cite{sfd}
The necessary conditions for this conclusion are basically the same as
those of the previous sections.
However, one must be more careful in avoiding a hot-electron effect
because the NEN is more sensitive to the electron temperature
than $G$ is.\cite{su,isqm,us}

Kane and Fisher\cite{KF} and Chamon et al.\cite{chamon}
studied the NEN of TL liquids {\it with barriers}.
Since their discussions relied on a perturbation expansion
which is good only when $G$ is small,
their results cannot apply to clean TL liquids without barriers.
On the other hand, our result is non-perturbative and thus is
valid for clean liquids (but is not applicable to dirty liquids).

The present new prediction, that the NEN is absent in clean
TL and Fermi liquids,
may be tested in careful experiments.

\section{Summary and Discussions}

We have shown that the two-terminal conductance $G$ of 1D interacting
electron systems is given by eq.\ (\ref{LF}) if the systems are
clean enough.
We have studied both Fermi liquids and TL liquids.
It is found that the physical mechanism leading to eq.\ (\ref{LF})
is basically the same in both cases:
$J$ and $\Delta \mu$ are
renormalized\cite{ren} by e-e interactions,
but the renormalization factors for $J$ and $\Delta \mu$
are the same, hence
their ratio ($=G$) is {\it not} renormalized at all.
We have also shown that the nonequilibrium noise is absent
in such a case.

The necessary conditions for these conclusions are the following.
(i) The 1D system is clean enough
and the temperature is low
enough (zero temperature has been assumed for simplicity),
so that scattering by impurities, defects or phonons
is negligible and e-e interactions are the only scattering mechanism.
(ii) The low-energy excitation spectrum of the 1D system
is gapless. (Although we have not stated this point explicitly,
it is clear that our discussions rely on the absence of a gap.)
(iii) The boundaries between
the 1D system and the reservoirs
are smooth and long ($>$ Fermi wavelength),
so that undesirable reflections at the boundaries are absent.
(iv) The reservoirs are large enough,
so that they remain at equilibrium even in the presence of
a finite current between the reservoirs through the 1D system.

When some of these conditions are not satisfied
the observed conductance may deviate from eq.\ (\ref{LF}).
We point out one example.
If dissipation (by, say, phonon emission) is nonnegligible,
the 1D system will lose any correlations over a distance $L_{\rm rlx}$,
where $L_{\rm rlx}$ is the
``maximal energy relaxation length",\cite{su,isqm,us}
which is generally longer than the simple dephasing length (over which
an energy correlation may be able to survive).\cite{su,isqm,us}
In such a case the 1D system of length $L$ ($> L_{\rm rlx}$) will behave as
a series of independent conductors of length $L_{\rm rlx}$.
One will then observe Ohm's law:
\begin{equation}
G_{obs} \simeq (L_{\rm rlx}/L)\times(s/2 \pi).
\end{equation}


\noindent
{\bf Note added in proof:}
We want to stress again that
our results are quite general. In fact,
the arguments in \S\ref{GF} and \S\ref{Fobs} for Fermi liquids
have not assumed any specific values of $f$'s, and
the arguments in \S\ref{GTL} and
\S\ref{TLobs} for TL liquids have merely assumed the forms of
$H$ and $J$, eqs.\ (\ref{H}) and (\ref{JTL}),
{\it only at the low-energy fixed point}.\cite{TL3,TL4,TL5}
Note that an {\it interacting} Fermi liquid, in which
quasi-particles are accompanied with the backflow,
might be realized in certain samples. Its conductance
has been exactly calculated for the
first time in this paper. In contrast,  previous
studies\cite{early1,early2,early3,early4,maslov,pono} analyzed
Fermi liquids only when interactions and the backflow are absent.
It is also worth noting that
the result (\ref{wrong}) of early theoretical
studies\cite{early1,early2,early3,early4}
can be reproduced
if we forget the renormalization of $\Delta \mu$.
This is consistent with Kawabata's conclusion\cite{kawabata} that
the neglect of renormalization of the electric field
leads to eq.\ (\ref{wrong}).

\section*{Acknowledgments}

The author thanks M. Ueda for discussions.
This work has been supported by
Grants-in-Aid for Scientific Research from the Sumitomo
Foundation and the
Ministry of Education, Science and Culture.


\begin{thebibliography}{99}
\bibitem{f}
Unlike in ref.\ \cite{noz}, we here separate the
trivial dependence on the
system size ($\propto 1/L$) from $f$.
\bibitem{sfd}
The absence of the NEN in
the strongly dissipative case can also be explained
by the shifted Fermi distribution.\cite{su,isqm}
\bibitem{ren}
By renormalization we here mean the fact that
the value of $J$ or $\Delta \mu$ deviates from
the free-electron value
for a given $q$ in the Fermi liquid case or for a given $N_\pm$ in
the TL liquid case.
On the other hand,  if we look at $J$ {\it for a given $\Delta \mu$}
then $J$ is {\it not} renormalized at all.
The term ``renormalization" should be used carefully.
\end{thebibliography}
\end{document}